\documentclass[11pt]{article}
\pdfoutput=1 

\usepackage{jheppub} 

\usepackage[T1]{fontenc} 

\newcommand{\fft}[2]{\frac{#1}{#2}}
\newcommand{\ft}[2]{{\textstyle\frac{#1}{#2}}}

\preprint{MCTP-15-23}
\title{\boldmath A Holographic $c$-Theorem for Schr\"odinger Spacetimes}

\author{James T. Liu}
\author{and Weishun Zhong}

\affiliation{Michigan Center for Theoretical Physics, Randall Laboratory of Physics,\\The University of Michigan, Ann Arbor, MI 48109--1040, USA}

\emailAdd{jimliu@umich.edu}
\emailAdd{wilsonzh@umich.edu}

\abstract{We prove a $c$-theorem for holographic renormalization group flows in a
Schr\"odinger spacetime that demonstrates that the effective radius $L(r)$ monotonically
decreases from the UV to the IR, where $r$ is the bulk radial coordinate.  This result
assumes that the bulk matter satisfies the null energy condition, but holds regardless of
the value of the critical exponent $z$.  We also construct several numerical examples in a
model where the Schr\"odinger background is realized by a massive vector coupled to a real scalar.
The full Schr\"odinger group is realized when $z=2$, and in this case it is possible to construct
solutions with constant effective $z(r)=2$ along the entire flow.}

\begin{document} 
\maketitle
\flushbottom

\section{Introduction}
\label{sec:intro}

In a relativistic conformal field theory, the Weyl anomaly $\langle T_\mu^\mu\rangle=\mathcal A$
signifies a breakdown of conformal invariance
at the quantum level, and plays an important role in the characterization of the theory.  This is
especially true in two dimensions, where the Cardy formula relates the central charge $c$ to the
degrees of freedom of the theory \cite{Cardy:1986ie}.  Moreover, the Zamolodchikov $c$-theorem
demonstrates that it is possible to define a $c$-function that is monotonic decreasing along renormalization
group flows from the UV to the IR \cite{Zamolodchikov:1986gt}.  These powerful statements have
seen recent generalizations to four and higher dimensions as well
\cite{DiPietro:2014bca,Ardehali:2014zba,Komargodski:2011vj}.

From a relativistic AdS/CFT point of view, the leading holographic Weyl anomaly is easily obtained from
the behavior of the on-shell action under rescaling of the boundary metric
\cite{Henningson:1998gx}.  For example, for AdS$_{d+1}$, the leading contribution to the $a$
central charge is
\begin{equation}
a=\fft{2d\pi^{d/2}}{\kappa^2(d/2)!^2}L^{d-1},
\label{eq:a-function}
\end{equation}
where $L$ is the AdS radius and $\kappa$ is the gravitational coupling.
While this is the result for pure, a holographic renormalization
group flow may be realized geometrically by turning on a relevant deformation and then solving the
equations of motion for radial evolution in the bulk.  In particular, the 
AdS metric in the Poincar\'e patch
\begin{equation}
ds_{d+1}^2=e^{2r/L}(-dt^2+d\vec x_{d-1}^2)+dr^2,
\end{equation}
has a natural domain wall generalization
\begin{equation}
ds_{d+1}^2=e^{2A(r)}(-dt^2+d\vec x_{d-1}^2)+dr^2.
\end{equation}
A flow between UV and IR fixed points is then given by the solution for $A(r)$ satisfying
$A\sim r/L_{\rm UV}$ as $r\to\infty$ and $A\sim r/L_{\rm IR}$ as $r\to-\infty$.  For such
a flow, we may define an $a$-function by replacing the constant AdS radius $L$ in
(\ref{eq:a-function}) by the effective radius $L(r)=1/A'(r)$.

In this context, the holographic $c$-theorem \cite{Alvarez:1998wr,Girardello:1998pd,Freedman:1999gp,Sahakian:1999bd} states that the effective AdS
radius $L(r)$  (and hence the $a$ function) is monotonic decreasing towards the IR.
For Einstein gravity in the bulk, this follows directly from the null energy condition
\begin{equation}
R_{\mu\nu}l^\mu l^\nu=\kappa^2T_{\mu\nu}l^\mu l^\nu\ge0,
\label{eq:nec}
\end{equation}
for all future-directed null vectors $l^\mu$.
In particular, choosing $l^\mu$ in the $t$--$r$ direction gives
\begin{equation}
R^r_r-R^t_t=-(d-1)A''\ge0\qquad\Rightarrow\qquad L'=-\fft{A''}{(A')^2}\ge0.
\label{eq:holoc}
\end{equation}
So long as we restrict to classical Einstein gravity in the bulk, the statement $L'\ge0$ is
completely general (as long as we impose the null energy condition), and moreover holds
in any spacetime dimension.

Given recent interest in non-relativistic holography, it is natural to ask whether a similar
$c$-theorem can be shown in the context of Lifshitz \cite{Taylor:2008tg,Nishida:2007pj}
or Schr\"odinger \cite{Son:2008ye,Balasubramanian:2008dm}
backgrounds.  (The Schr\"odinger case has been considered previously in \cite{Myers:2010tj}.)
In the Lifshitz case, however, it was shown in \cite{Liu:2012wf} that the null
energy condition does not constrain the effective
radius $L(r)$, so that it is not necessarily monotonic along the flow and can actually increase
toward the IR.  In particular, starting from the Lifshitz metric
\begin{equation}
ds_{d+2}^2=-e^{2zr/L}dt^2+e^{2r/L}d\vec x_d^2+dr^2,
\end{equation}
with critical exponent $z$, we may construct a domain wall solution of the form
\begin{equation}
ds_{d+2}^2=-e^{2A(r)}dt^2+e^{2B(r)}d\vec x_d^2+dr^2.
\label{eq:ABlif}
\end{equation}
In order to study Lifshitz flows, it is useful to define the flow functions
\begin{equation}
L(r)=\fft1{B'(r)},\qquad z(r)=\fft{A'(r)}{B'(r)}.
\label{eq:ff}
\end{equation}
When applying the null energy condition, we may choose a null vector either along $t$--$x$ or
along $t$--$r$.  Assuming $z\ge1$, we are led to two inequalities \cite{Liu:2012wf}
\begin{equation}
L'\ge-(z-1),\qquad
z'\ge-(z-1)(d+2z-1)/L.
\label{eq:Lifineq}
\end{equation}
However, since the right-hand sides of both expressions are negative for $z>1$, neither inequality
leads to monotonicity of the respective flow functions.

The Lifshitz flow reduces to the relativistic case in the limit $A=B$ (or equivalently $z=1$).
In this limit, the first inequality in (\ref{eq:Lifineq}) reduces to $L'\ge0$, which reproduces the
relativistic $c$-theorem, while the second becomes trivial.  This suggests that additional symmetry
beyond that of the Lifshitz metric is required to obtain monotonic behavior of the flow functions.
One natural possibility is to consider Schr\"odinger holography \cite{Son:2008ye,Balasubramanian:2008dm}
 where the metric can be
written in the form
\begin{equation}
ds_{d+3}^2=-e^{2zr/L}dt^2+e^{2r/L}(2\,dt\,d\xi+d\vec x_d^2)+dr^2.
\label{eq:Sch}
\end{equation}
In addition to the radial direction $r$, the Schr\"odinger metric also includes $\xi$ which is the
coordinate conjugate to conserved particle number.  In this paper, we show that, in contrast with the
Lifshitz case, the null energy condition and the Einstein equation is sufficient to demonstrate
the monotonicity of the effective radius $L(r)$%
\footnote{Apparently scalars with sufficiently negative $m^2$ can exhibit limit cycle behavior in
$z=2$ Schr\"odinger spacetimes \cite{Moroz:2009kv}.  It would be interesting to see how that ties
in with monotonicity of $L(r)$.}.

In addition to proving that $L(r)$ is monotonic in Schr\"odinger backgrounds, we investigate
holographic RG flows in a simple model where the bulk metric is supported by a massive
vector coupled to a real scalar.  By choosing appropriate potentials, we can realize flows with
$z_{UV}=z_{IR}$ as well as with $z_{UV}\ne z_{IR}$.  While $L(r)$ is indeed monotonic along
flows, we find it easy to construct numerical flows where the effective $z(r)$ fails to be monotonic.
Additional symmetry arises for $z=2$ Schr\"odinger, and we see that in this case a judicious
choice of potentials allows us to construct solutions where $z=2$ is constant along the entire flow.
Holographic flows from $z=1$ AdS to $z=2$ Schr\"odinger have been constructed previously
in \cite{Ishii:2011dt} in the context of a consistent truncations of IIB supergravity and M-theory.

This paper is organized as follows.  In section~\ref{section 2}, we study the consequences of the
null energy condition and prove a Schr\"odinger $c$-theorem showing that $L'\ge0$.
Although the full Schr\"odinger symmetry is only realized for $z=2$, monotonicity of $L$
holds for arbitrary $z\ge1$.  In section~\ref{section 3}, we study numerical flows in a simple
massive vector coupled to scalar model.  Finally, we conclude in section~\ref{sec:disc}
with a brief mention of the connection between the effective radius $L(r)$ and non-relativistic scale
anomalies.  Although the null energy condition does not lead to monotonicity of $L(r)$ in
Lifshitz holography, we consider the possibility of using the weak energy condition to derive
a corresponding Lifshitz $c$-theorem in the appendix.

\section{Holographic c-theorem in Schr\"odinger spacetime}
\label{section 2}

In order to describe a Schr\"odinger flow, we generalize the metric (\ref{eq:Sch}) away from fixed
points by taking
\begin{equation}
\label{eq:2.3}
ds_{d+3}^2=-e^{2A(r)}dt^2+e^{2B(r)}(2\,dt\,d\xi+d\vec x_d^2)+dr^2.
\end{equation}
Note that $\partial/\partial\xi$ remains a null Killing vector everywhere along the flow.
Following \cite{Liu:2012wf}, we use the same definitions of the flow functions (\ref{eq:ff}) as was
used in the Lifshitz case.  Both $L(r)$ and $z(r)$ approach constants
$L_{\rm UV}$, $z_{\rm UV}$ and $L_{\rm IR}$, $z_{\rm IR}$ at the fixed points of the flow.

\subsection{Applying the null energy condition}

In order to apply the null energy condition, we first compute the Ricci tensor for the metric
(\ref{eq:2.3}) in terms of the flow functions $L(r)$ and $z(r)$.  The result is
\begin{equation}\begin{split}
R_{tt} &= -{\frac{g_{tt}}{L^2}}(2z^2+(d-2)z+2+z'L-zL'),\\
R_{ij} &=-\frac{g_{ij}}{L^2}(d+2-L'),\\
R_{t\xi}&=-\frac{g_{t\xi}}{L^2}(d+2-L'),\\
R_{rr} &=-\fft1{L^2}(d+2)(1-L').
\end{split}\end{equation}
We now consider the null energy condition (\ref{eq:nec}).  In contrast with the relativistic
case, the condition depends on the choice of the null vector field, and we find
\begin{equation}
(d+2z)(z-1)+z'L-(z-1)L'+\alpha (d+1)L'\ge0,
\end{equation}
where
\begin{equation}
\alpha=\left|\fft{g_{rr}l^rl^r}{g_{tt}l^tl^t}\right|\ge0.
\end{equation}
The value of $\alpha$ depends on the null vector field, and ranges from $0$ (e.g.\ for a
null vector in the $t$--$x$ direction) to $\infty$, which is obtained in the limit when
$l^\mu$ points mostly along the $\xi$ direction:
\begin{equation}
l^\mu\fft\partial{\partial x^\mu}=\epsilon e^{-A}\fft\partial{\partial t}-\fft{e^{A-2B}}{2\epsilon}
\fft\partial{\partial \xi}+\sqrt{1+\epsilon^2}\fft\partial{\partial r}\qquad\mbox{as}\quad\epsilon\to0.
\label{eq:nulllim}
\end{equation}

The limiting values of $\alpha$ give rise to two inequalities
\begin{equation}
L'\ge0,\qquad (z-1)L'\le(d+2z)(z-1)+z'L.
\label{eq:twoineq}
\end{equation}
The first inequality demonstrates that the effective radius $L(r)$ is monotonically increasing
towards the UV.  This may be viewed as a non-relativistic generalization of the holographic
$c$-theorem, (\ref{eq:holoc}).  It is worth noting that this inequality arises in the limiting case
when the null vector $l^\mu$ is directed along $\xi$--$r$, as in (\ref{eq:nulllim}).  This singles
out the metric function $B(r)$, and hence isolates the effective radius $L(r)=1/B'(r)$.  In particular,
this choice is unavailable in the Lifshitz case, where the metric takes the form (\ref{eq:ABlif}),
and where the null vector must necessarily include the $A(r)$ function.  This is the underlying
reason for the lack of monotonicity of the effective radius in Lifshitz flows \cite{Liu:2012wf}.

If we restrict to the case $z>1$, then (\ref{eq:twoineq}) also gives an upper bound on $L'$
\begin{equation}
0\le L'\le d+2z+\fft{z'L}{z-1}.
\label{eq:ctheorem}
\end{equation}
Combining both inequalities then yields a lower bound on $z'$
\begin{equation}
z'\ge-(z-1)(d+2z)/L.
\end{equation}
This is similar to the bound (\ref{eq:Lifineq}) obtained for Lifshitz flows, except that the effective
dimension $d+2z$ is increased by one (corresponding to the addition of the $\xi$ coordinate
in the Schr\"odinger bulk).

For the relativistic case, $z=1$, the second inequality in (\ref{eq:twoineq}) becomes trivial,
and the upper limit on $L'$ is removed.  As a result, we recover the relativistic $c$-theorem
\cite{Alvarez:1998wr,Girardello:1998pd,Freedman:1999gp,Sahakian:1999bd}.
For $z<1$, both inequalities in (\ref{eq:twoineq}) provide lower bounds on $L'$.  However,
note that such flows cannot have any fixed points, as setting $L'=z'=0$ in (\ref{eq:twoineq})
yields $(z-1)(d+2z)\ge0$, so that $z\ge1$ at fixed points.  (Here we ignore the possibility
that $z\le-d/2$.)

\section{Schr\"odinger flows in a phenomenological model}
\label{section 3}

We now turn to some examples of Schr\"odinger flows between UV and IR fixed points.
Our starting point is the massive vector (or equivalently abelian Higgs in its broken phase)
model with action \cite{Son:2008ye,Balasubramanian:2008dm}
\begin{equation}
\label{eq:3.1}
S=\int d^{d+3}x\sqrt{-g}(R-2\Lambda-\frac{1}{4}F_{\mu\nu}F^{\mu\nu}-\frac{1}{2}m^2A_\mu A^\mu).
\end{equation}
This admits a solution where the Schr\"odinger metric (\ref{eq:Sch}) is supported by the vector
field
\begin{equation}
A=\sqrt{\fft{2(z-1)}z}e^{zr/L}dt.
\end{equation}
The constants $z$ and $L$ are related to the theory parameters $m^2$ and $\Lambda$ according to
\begin{equation}
\label{eq:3.2}
m^2=\frac{z(z+d)}{L^2},\qquad
\Lambda=-\frac{(d+1)(d+2)}{2L^2}.
\end{equation}
In particular, once $m^2$ and $\Lambda$ are chosen, the scaling behavior is uniquely determined.
This is in contrast with the Lifshitz case \cite{Braviner:2011kz}, where it is possible to have two fixed points (and
hence flows between fixed points) for the same theory parameters.

In order to construct flows between different Schr\"odinger fixed points, we generalize (\ref{eq:3.1})
to allow for dynamical $m^2$ and $\Lambda$ by adding a real scalar field
\begin{equation}
\label{eq:3.3}
S=\int d^{d+3}x\sqrt{-g}(R-2V(\phi)-\frac{1}{2}\partial_\mu \phi \partial^\mu \phi-\frac{1}{4}F_{\mu\nu}F^{\mu\nu}-\frac{1}{2}W(\phi)A_\mu A^\mu).
\end{equation}
This model was previously considered in \cite{Liu:2012wf} in the Lifshitz context.
To proceed, we use the domain wall ansatz \eqref{eq:2.3} and take matter fields to be
\begin{equation}
A =H(r)e^{A(r)}dt\,,
\qquad
\phi=\phi(r)\,.
\end{equation}
The scalar and vector equations of motion are
\begin{equation}\begin{split}
0 &= \phi'' + (d+2)\phi'B' - 2\partial_\phi V,\\
0 &= A''H + A'^2 H+ 2A'H'+dB'(H'+A'H)+H''-WH,
\end{split}\end{equation}
and the Einstein equations give rise to
\begin{equation}\begin{split}
0 &= A''-B''+2A'^2 +(d-2)A'B'-dB'^2 -\frac{1}{2}[(H'+A'H)^2+WH^2],\\
0 &= (d+1)B''+\frac{1}{2}\phi'^2,
\end{split}\end{equation}
along with the constraint equation
\begin{equation}
0 = \frac{(d+1)(d+2)}{2}B'^2+V-\frac{1}{4}\phi'^2.
\end{equation}

The above equations of motion can be rewritten in terms of the flow functions $L(r)$ and $z(r)$
defined in \eqref{eq:ff}.  The result is
\begin{equation}\begin{split}
\label{eq:eom}
0 &= \frac{z'L-L'z}{L^2}H+\frac{z^2}{L^2}H+\frac{2z}{L}H'+\frac{d}{L}(H'+\frac{z}{L}H)+H''-WH,
\\
0 &= \phi'' + \frac{d+2}{L}\phi'-2\partial_\phi V,
\\
0 &= \frac{1}{L^2}[z'L+(1-z)L'+(d+2z)(z-1)]-\frac{1}{2}[(H'+\frac{z}{L}H)^2+WH^2],
\\
0 &= -\frac{d+1}{L^2}L' + \frac{1}{2}\phi'^2.
\end{split}\end{equation}
In addition, the constraint equation becomes
\begin{equation}
\label{eq:conseqn}
0 = \frac{(d+1)(d+2)}{2L^2}+V-\frac{1}{4}\phi'^2.
\end{equation}
It is evident that the last equation in \eqref{eq:eom} immediately gives rise to the restriction $L'\ge0$,
in agreement with the lower bound from the $c$-theorem, (\ref{eq:ctheorem}).
Note, however, that the null energy condition, as a constraint on the stress-energy tensor, requires that
\begin{equation}
\label{eq:3.9}
W(\phi) \geq -\frac{z}{L^2}.
\end{equation}
Any $W(\phi)$ we choose must satisfy this requirement everywhere along the flow.

\subsection{Fixed points}
Before turning to flows, we first examine the fixed point behavior of this system.  Substituting the
constant values
\begin{equation}
L(r) =L_0\,,
\qquad
z(r) =z_0\,,
\qquad
\phi(r) =\phi_0,
\end{equation}
into the equations of motion \eqref{eq:eom} and \eqref{eq:conseqn} gives
\begin{equation}\begin{split}
\label{eq:fpcond}
W(\phi_0) &=\frac{z_0(z_0+d)}{L_0^2},
\\
V(\phi_0) &=-\frac{(d+1)(d+2)}{2L_0^2},
\\
\partial_\phi V(\phi_0) &=0,
\\
H_0 &= \sqrt{\frac{2(z_0-1)}{z_0}},
\end{split}\end{equation}
which, not surprisingly, agrees with \eqref{eq:3.2}.

\subsection{Linearized analysis}

As a guide for constructing flows, we now proceed to linearize the equations of motion \eqref{eq:eom}
in the vicinity of a fixed point according to the following recipe
\begin{equation}
L =L_0+\epsilon\hat{L},\qquad     z =z_0+\epsilon\hat{z} ,   \qquad
\phi =\phi_0+\epsilon\hat{\phi},\qquad        H =H_0+\epsilon\hat{H}.
\end{equation}
Although the first two equations in (\ref{eq:eom}) are second order in $H$ and $\phi$, they may be
rewritten as a set of first order equations by introducing $\hat H'$ and $\hat\phi'$ as independent
functions.  For $\epsilon \ll 1$, we end up with a system of first order linear differential equations
\begin{equation}
\mathcal{V'=MV},
\end{equation}
where
\begin{equation}
\mathcal{V}=(\hat{L}\,,\hat{z}\,,\hat{\phi}\,,\hat{\phi'}\,,\hat{H}\,,\hat{H'})^T,
\end{equation}
and
\begin{equation}
\mathcal{M}= 
\begin{pmatrix}
  0 & 0 & 0 & 0 & 0 & 0 \\
  \frac{2(z_0-1)(d+z_0)}{L_0^2} & -\frac{ d+ 2z_0 }{L_0} & \frac{L_0 W_1(z_0-1)}{z_0} & 0 &\fft{z_0 H_0(d+2z_0)}{L_0} & z_0 H_0 \\
  0 & 0 & 0 & 1 & 0 & 0 \\
  0 & 0 & 2V_2&-\frac{d+2}{L_0}  & 0 & 0 \\
  0 & 0 & 0 & 0 & 0 & 1 \\
  \frac{2H_0 (d+z_0)}{L_0^3}& 0 & \fft{H_0 W_1}{z_0} & 0 &
 -\fft{2(z_0-1)(d+2z_0)}{L_0^2} &- \fft{d+4z_0-2}{L_0}
 \end{pmatrix}.
\end{equation}
Note that we have expand the potential $V(\phi)$ and effective mass term $W(\phi)$ around the fixed point $\phi=\phi_0$
\begin{equation}
\begin{split}
V(\phi)&=V_0 + V_1(\phi-\phi_0) + \frac{1}{2}V_2 (\phi-\phi_0)^2 + \dotsb,
\\
W(\phi)&=W_0 + W_1(\phi-\phi_0) + \frac{1}{2}W_2 (\phi-\phi_0)^2 + \dotsb.
\end{split}
\end{equation}

The solution to this system of first order equations may be written in the general form
\begin{equation}
\mathcal{V}(r)=\displaystyle\sum_{i}\mathcal{V}_i e^{\lambda_ir},
\end{equation}
where $\{\lambda_i\}$ are the eigenvalues of $\mathcal M$ and $\{\mathcal{V}_i\}$ are the corresponding eigenvectors.  Taking $r\to\infty$ to be the UV, the negative eigenvalues $\lambda<0$ correspond to
relevant deformations, as they correspond to flows away from the fixed point as $r$ is decreased towards
the IR.  To have a stable flow from the UV to the IR, we must move away from the UV in a relevant
direction ($\lambda_{\rm UV}<0$) and approach the IR along an irrelevant direction ($\lambda_{\rm IR}>0$).

The eigenvalues of the system can be determined by solving the secular equation.  
There is one marginal deformation with
\begin{equation}
\lambda_1=0,\qquad\mathcal V_1=\left(L_0(z_0-1)\fft{d+2z_0}{d+z_0},2z_0(z_0-1),0,0,H_0,0\right),
\label{eq:marg}
\end{equation}
corresponding to a shift in $z_0$ and $L_0$ leaving $W(\phi_0)$ unchanged in (\ref{eq:fpcond}).
Note, however, that this shift will affect the value of $V(\phi_0)$, so it is actually removed
by the constraint equation (\ref{eq:conseqn}).  We also find a relevant deformation with
\begin{equation}
\lambda_2=-\fft{d+2z_0}{L_0},\qquad\mathcal V_2=\left(0,1,0,0,0,0\right).
\end{equation}
This corresponds to a flow in $z$ with fixed $L$, at least initially along the flow.

The remaining four eigenvalues come in two pairs.  The first pair is
\begin{equation}\begin{split}
\lambda_3&=-\fft{d+2z_0}{L_0},\kern2.8em\mathcal V_3=\left(0,0,0,0,L_0,-(d+2z_0)\right),\\
\lambda_4&=-\fft{2(z_0-1)}{L_0},\qquad\mathcal V_4=\left(0,z_0H_0L_0,0,0,L_0,-2(z_0-1)\right).
\end{split}\end{equation}
Both of these deformations are relevant.  Moreover, since the corresponding eigenvectors are
involve $\hat H$ and $\hat H'$ but not $\hat\phi$ nor $\hat\phi'$, we denote these flows as
`vector field driven'.  In contrast, the final two flows are `scalar field driven', and have eigenvalues
\begin{equation}\begin{split}
\lambda_{5,6}=-\fft{\Delta_\pm}{L_0},\qquad&\mathcal V_{5,6}
=\left(0,0,1,-\fft{\Delta_\pm}{L_0},0,0\right)\\
&\kern3em
+\fft{W_1L_0}{2z_0\Xi_\pm}\left(0,-L_0(z_0-1)(\Delta_\pm-2z_0),0,0,H_0L_0,-H_0\Delta_\pm\right),
\qquad
\label{eq:irrel}
\end{split}\end{equation}
where
\begin{equation}\begin{split}
\Delta_\pm&=\fft{d+2\pm\sqrt{(d+2)^2+8V_2L_0^2}}2,\\
\Xi_\pm&=V_2L_0^2+(z_0-1)\left(2(z_0-1)\mp\sqrt{(d+2)^2+8V_2L_0^2}\right).
\end{split}\end{equation}
Note that the eigenvectors simplify considerably when $W_1=0$, in which case the linear coupling
between $\phi$ and $A_\mu^2$ vanishes at the fixed point.  The deformation corresponding to $\Delta_+$
is always relevant, while the deformation corresponding to $\Delta_-$ is irrelevant for $V_2>0$,
marginal for $V_2=0$ and relevant for $-(d+2)^2/8L_0^2\le V_2<0$.

\subsection{Numerical solution}

We construct flows by solving the equations of motion (\ref{eq:eom}) using the shooting method.
Ignoring the marginal deformation (\ref{eq:marg}) which takes us out of the vacuum imposed by
(\ref{eq:conseqn}), at any fixed point there are four relevant deformations and a fifth deformation
that is either relevant or irrelevant depending on the second derivative of the potential.  It is thus
natural to shoot from the IR fixed point to the UV by moving along the single irrelevant direction.

We must, of course, specify the potential $V(\phi)$ and scalar coupling $W(\phi)$ before proceeding.
Since we aim to flow between two fixed points, we need a potential with at least two critical points.
For the simplest case, we take cubic functions
\begin{equation}
\begin{split}
\label{eq:3.24}
V(\phi)&= V_0 + V_1 \phi+ \ft12V_2\phi^2 + \ft16V_3\phi^3,
\\
W(\phi)&= W_0 + W_1 \phi+ \ft12W_2\phi^2 + \ft16W_3\phi^3.
\end{split}
\end{equation} 
Assuming a flow from $\phi=0$ in the IR to $\phi=\phi_0$ in the UV, and taking the first derivative of
$W(\phi)$ to vanish at fixed points, the fixed point conditions (\ref{eq:fpcond}) give rise to the unique
set of coefficients
\begin{equation}
\label{eq:3.25}
\begin{split}
V_0 &=-\frac{(d+1)(d+2)}{2L_{IR}^2},
\\
V_1 &=0,
\\
V_2 \phi_0^2 &=-3(d+1)(d+2)\left(\frac{1}{L_{UV}^2}-\frac{1}{L_{IR}^2}\right),
\\
V_3 \phi_0^3 &=6(d+1)(d+2)\left(\frac{1}{L_{UV}^2}-\frac{1}{L_{IR}^2}\right),
\end{split}
\end{equation}
and
\begin{equation}
\label{eq:3.26}
\begin{split} 
W_0 &=\frac{z_{IR}(d+z_{IR})}{L_{IR}^2},
\\
W_1 &=0,
\\
W_2 \phi_0^2 &=6\left(\frac{z_{UV}(d+z_{UV})}{L_{UV}^2}-\frac{z_{IR}(d+z_{IR})}{L_{IR}^2}\right),
\\
W_3 \phi_0^3 &=12\left(\frac{z_{IR}(d+z_{IR})}{L_{IR}^2}-\frac{z_{UV}(d+z_{UV})}{L_{UV}^2}\right).
\end{split}
\end{equation}
Note that the cubic form of $W(\phi)$ is unbounded from below, and will not satisfy the null
energy constraint \eqref{eq:3.9} for all values of the field $\phi$.  However, so long as \eqref{eq:3.9}
is satisfied everywhere along the flow, then the null energy condition will continue to hold for
the classical domain wall solution.  We verify that this is indeed the case for the numerical
solutions constructed below.

For the numerical solution, we set $\phi_0=1$ and start at the IR fixed point specified by
$(L_{IR},z_{IR},\phi_{IR},H_{IR})$ with $\phi_{IR}=0$ and
\begin{equation}
H_{IR} = \sqrt{\frac{2(z_{IR}-1)}{z_{IR}}}.
\end{equation}
We then move slightly away from the fixed point along the $\lambda_6$ direction in (\ref{eq:irrel}).
As a result, this flow is inherently scalar field driven.  In order
to ensure that this is an irrelevant direction, we must take $V_2>0$.  In this case, the expression
for $V_2$ in (\ref{eq:3.25}) immediately demands $L_{UV}>L_{IR}$.  (Although this is clearly compatible
with (\ref{eq:ctheorem}), it is by no means a proof of the $c$-theorem, as the $c$-theorem is a general
result, while here we are only working in a particular toy model.)

\subsubsection{Schr\"odinger flow with constant $z=2$}

\begin{figure}[t]
\begin{center}
\includegraphics{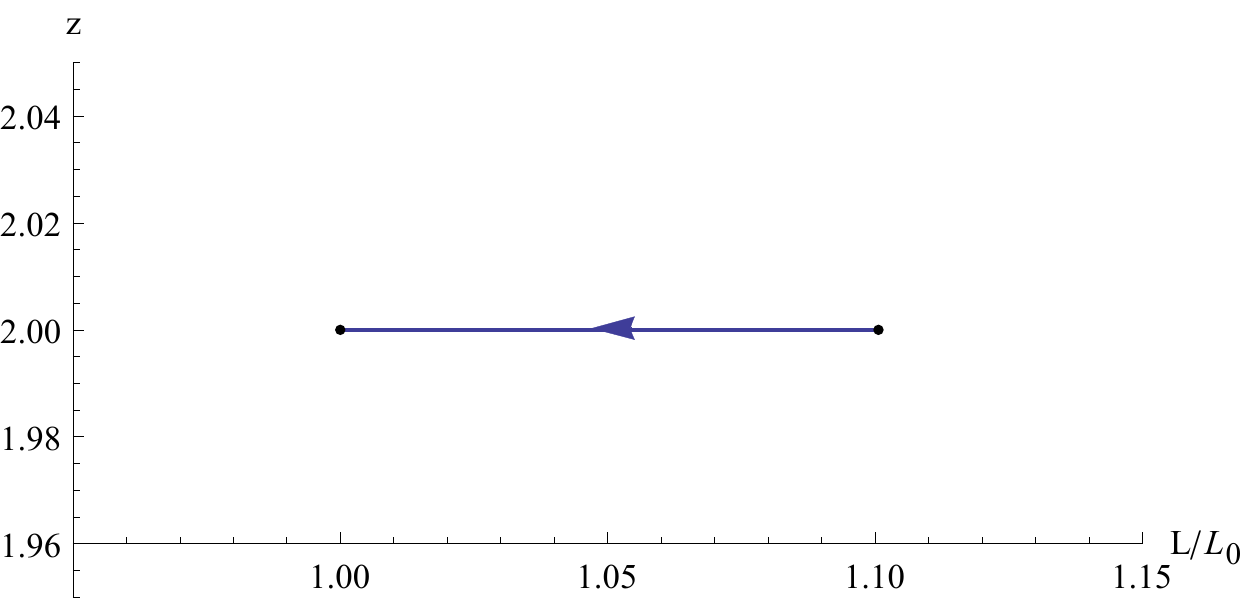}
\end{center}
\caption{\label{fig:Sch2} A solution with constant $z=2$ everywhere during the flow. The fixed point parameters are given by \eqref{eq:3.28} along with $d=3$ and $\phi_0=1$. The arrow indicates the flow direction from UV to IR.}
\end{figure}

Since the full Schr\"odinger symmetry is only realized for $z=2$, we first consider a flow with
$z_{UV}=z_{IR}=2$.  We take, as an example
\begin{equation}
\label{eq:3.28}
\begin{split}
(L_{UV},\;z_{UV}) &= (11L_0/10,\;2),\\
(L_{IR},\;z_{IR}) &= (L_0,\;2).
\end{split}
\end{equation} 
The numerical solution for the flow in the $z$--$L$ plane is shown in Fig.~\ref{fig:Sch2}.  As is
evident, the solution has constant effective $z(r)=2$ throughout the flow, even though this was
not implemented as a constraint in the massive vector coupled to scalar model of (\ref{eq:3.3}).
Moreover, the solution maintains $H(r)=1$, so that the vector field is of the form $A^\mu\partial_\mu
=\partial_\xi$.

As far as we have investigated, no other solutions with $z\ne2$ at the fixed points
have constant $z(r)$ along the flow.  This suggests that the additional Schr\"odinger symmetry
for $z=2$ allows for consistent flows with constant $z$.  In particular, imposing $z(r)=2$ and
$H(r)=1$ reduces the system of equations (\ref{eq:eom}) into four equations for two unknowns,
$L(r)$ and $\phi(r)$.  Since this system is over-constrained, some additional symmetry is needed
for consistency.  In this case, the key symmetry is the realization
that $A^\mu\partial_\mu$ is a null Killing vector for this $z=2$ flow.  The combination of
the Maxwell and Killing equations then give the constraint
\begin{equation}
WA_\nu=\nabla^\mu F_{\mu\nu}=-2\nabla^\mu\nabla_\nu A_\mu=-2\nabla_\nu\nabla^\mu A_\mu
-2R_\nu{}^\lambda A_\lambda.
\end{equation}
Since the massive vector is divergence-free by its equation of motion, we are left with
\begin{equation}
R_{\mu\nu}A^\nu = -\frac{1}{2}W(\phi)A_\mu.
\label{eq:Acnstr}
\end{equation}
On the other hand, contraction of $A^\lambda$ with the Einstein equation
\begin{equation}
R_{\mu\nu}=\ft12\partial_\mu\phi\partial_\nu\phi+\ft12(F_{\mu\lambda}F_\nu{}^\lambda-\ft1{2(d+1)}g_{\mu\nu}F^2)+\ft2{d+1}g_{\mu\nu}V+\ft12A_\mu A_\nu,
\end{equation}
gives
\begin{equation}
R_{\mu\nu}A^\nu=\fft2{d+1}V(\phi)A_\mu,
\label{eq:Aeins}
\end{equation}
provided $A^\mu$ is a null Killing vector and $A^\mu\partial_\mu\phi=0$.  Combining 
(\ref{eq:Acnstr}) with (\ref{eq:Aeins}) then gives the condition
\begin{equation}
\label{eq:3.29}
V(\phi) = -\frac{d+1}{4}W(\phi),
\end{equation}
which indeed holds for $z_{IR}=z_{UV}=2$ in the cubic potential (\ref{eq:3.24}).

The relation (\ref{eq:3.29}) is a necessary condition for $A^\mu$ to be a null Killing vector.  However,
it only removes one redundancy in the equations of motion.  The second redundancy comes from
comparing the Maxwell equation in the first line of (\ref{eq:eom}) with the combination of $ii$ and
$tt$ Einstein equations in the third line of (\ref{eq:eom}).  Setting $H=1$ in these two equations gives
\begin{equation}
\begin{split}
0 &= z'L - zL' + z(z+d) - WL^2,
\\
0 &= 2z'L - zL'+ z(z+d) - WL^2 + (z-2)(2z+d-L').
\end{split}
\end{equation}
These equations are redundant when $z(r)=2$, and we are left with a relatively simple system
\begin{equation}
\begin{split}
0 &= \phi'' + \frac{d+2}{L}\phi'- 2\partial_\phi V,
\\
0 &= L' - (d+2) - \frac{2L^2}{d+1}V.
\end{split}
\end{equation}
for the two functions $\phi(r)$ and $L(r)$. ( Alternatively, the scalar equation can be replaced by
the constraint \eqref{eq:conseqn}.)

What we have shown is that, when the potential satisfies \eqref{eq:3.29}, the massive vector coupled
to scalar model admits flows with $z=2$ and $H=1$ along the entire flow.  Of course, we can also
ask what happens when this constraint is not satisfied.  As we now show, while it is still possible to
flow from $z_{UV}=2$ to $z_{IR}=2$, the effective $z(r)$ will not be constant during the flow,
and neither will $H(r)$.

\subsubsection{Schr\"odinger flow with $z_{UV}=z_{IR}=2$, but changing $z$ in between}

Since the potential relation \eqref{eq:3.29} provides an additional symmetry allowing for constant
$z=2$ flows, we may break this symmetry by adding another term to $V(\phi)$ in \eqref{eq:3.24}.
In particular, we may add a quartic term to $V(\phi)$, while maintaining a cubic $W(\phi)$.  One
way to do this without affecting the UV and IR fixed point parameters is to add a term of the form
\begin{equation}
\label{eq:3.31}
V(\phi)= V_0 + V_1 \phi+ \ft12V_2\phi^2 + \ft16V_3\phi^3 + \ft1{24}V_4\phi^2(\phi-\phi_0)^2,
\end{equation}
where $V_4>0$, but is otherwise unconstrained.
Since the flow is engineered to go from $\phi=0$ in the IR to $\phi=\phi_0$ in the UV, the additional
term and its first derivative vanishes at the endpoints of the flow, thus ensuring that the fixed point
data in (\ref{eq:3.25}) remains unchanged.

\begin{figure}[t]
\begin{center}
\includegraphics{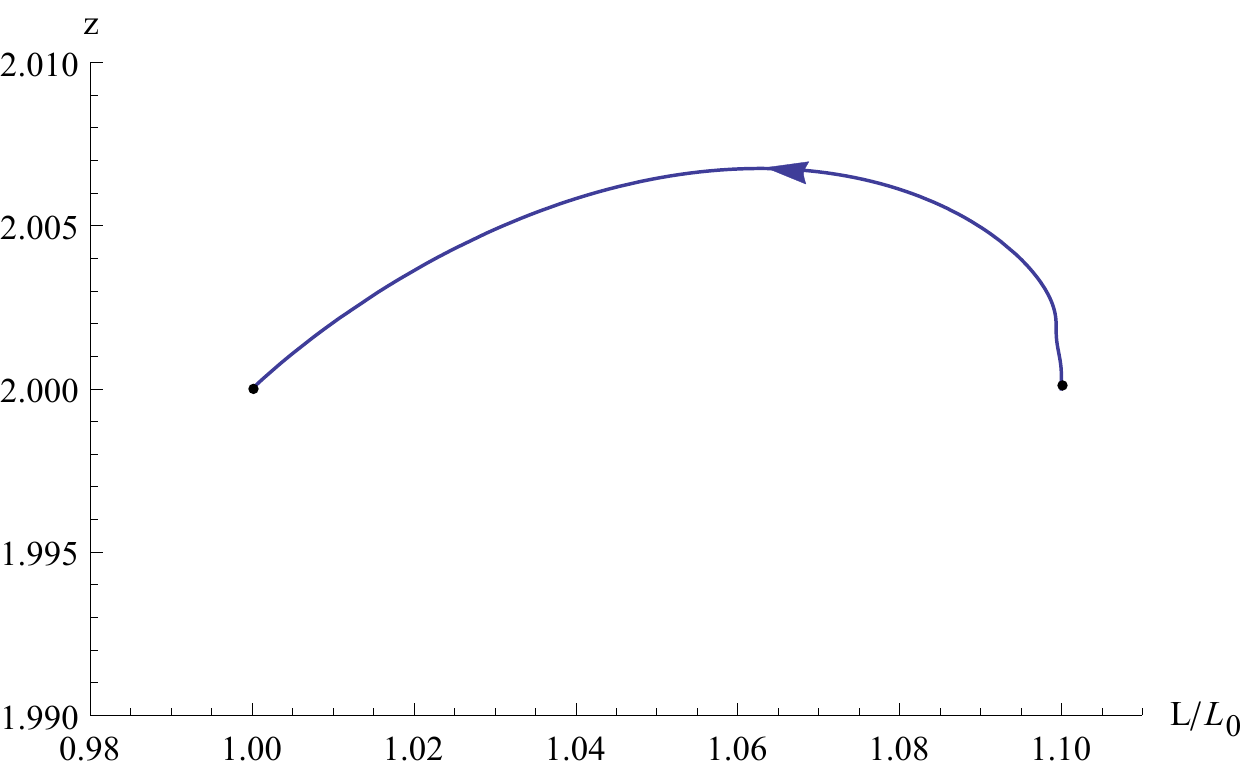}
\end{center}
\caption{\label{fig:2} A solution with $z_{UV}=z_{IR}=2$ using the potential (\ref{eq:3.31}).
The fixed point parameters are given by\eqref{eq:3.28} along with $d=3$, $V_4=24/\phi_0^2$
and $\phi_0=1$. The arrow indicates the flow direction from UV to IR.}
\end{figure}

As an example of a flow with non-constant $z(r)$, we choose the same fixed point parameters
(\ref{eq:3.28}) and take $V_4=24/\phi_0^2$.  The numerical flow is shown in Fig.~\ref{fig:2}.
Although $z(r)$ is no longer a constant along the flow, it starts and ends at the expected $z=2$
fixed points.  This shows explicitly that, while $L(r)$ remains monotonic decreasing towards the IR
(as it must by the null energy condition), $z(r)$ is certainly not monotonic.

\subsubsection{Schr\"odinger flow between different $z_{UV}$ and $z_{IR}$}

\begin{figure}[t]
\begin{center}
\includegraphics[scale=0.95]{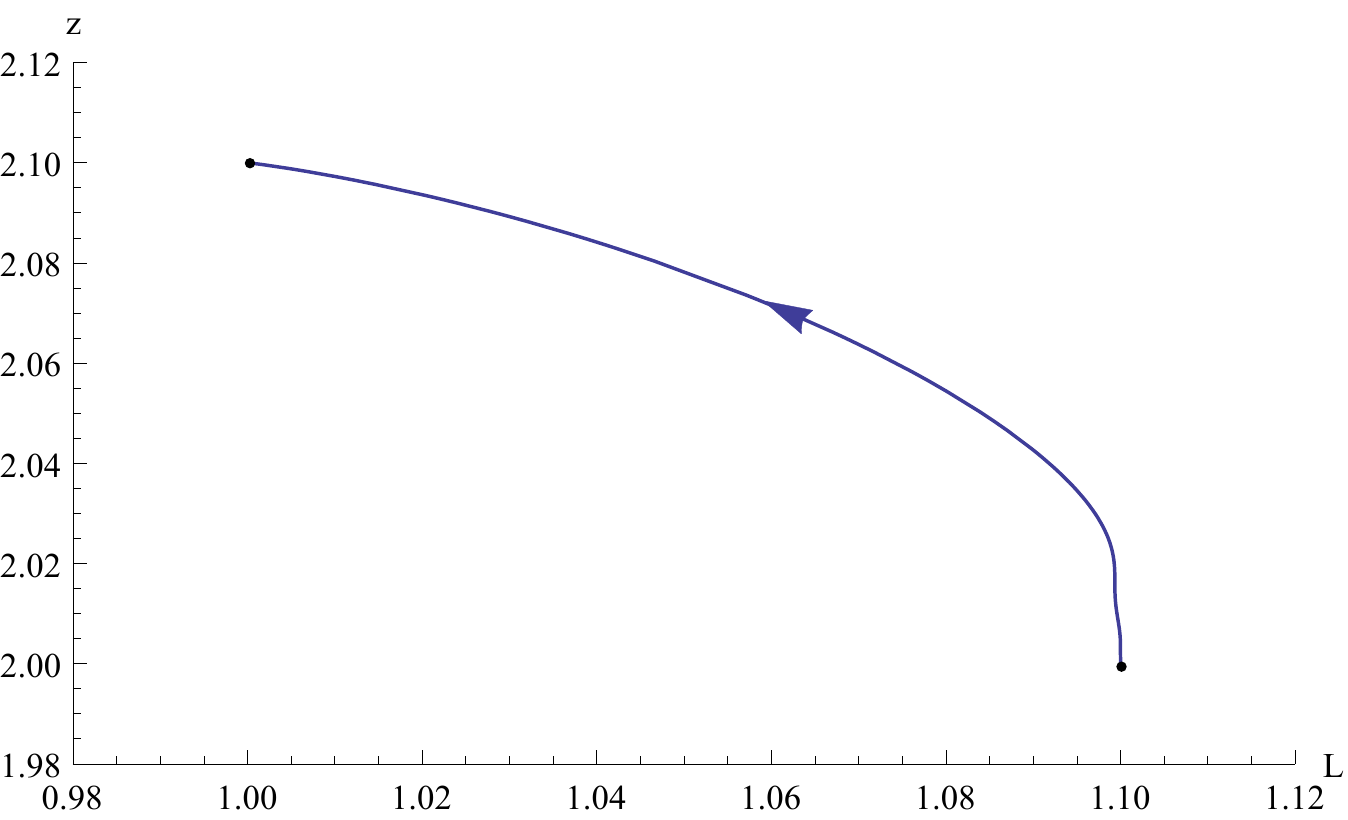}
\end{center}
\caption{\label{fig:3} A solution flowing from $z_{UV}=2$ to $z_{IR}=2.1$. The fixed point parameters are given by \eqref{eq:3.28} along with $d=3$, $V_4=24/\phi_0^2$ and $\phi_0=1$. The arrow indicates the flow direction from UV to IR.}
\end{figure}

The final example we consider is a flow with different fixed point $z$ values. We use the quartic
$V(\phi)$ in \eqref{eq:3.31} with $V_4=24/\phi_0^2$ along with the fixed point parameters
\begin{equation}
\label{eq:3.33}
\begin{split}
(L_{IR},\;z_{IR}) &= (L_0,\;21/10),
\\
(L_{UV},\;z_{UV}) &= (11L_0/10,\;2).
\end{split}
\end{equation}
The numerical solution is shown in Fig~\ref{fig:3}.
This solution also exhibits monotonicity in $L$ toward the IR. However, it is worth noticing that this does not agree with the result in the appendix of \cite{Myers:2010tj}, which claims that $L_{UV} > L_{IR}$ leads to $z_{UV} \geq z_{IR}$ in Schr\"odinger spacetimes.

\section{Discussion}\label{sec:disc}

Our formulation of a Schr\"odinger $c$-theorem is given in terms of the effective radius $L(r)$.
In the relativistic case, the AdS radius is directly related to the $a$ central charge according to
\eqref{eq:a-function}.  Hence monotonicity of $L(r)$ is equivalent to monotonicity of the corresponding
$a(r)$ function.  We would naturally like to make a similar connection between the effective
radius and the scale anomaly in the non-relativistic case.

The non-relativistic version of the Weyl anomaly is the quantum breakdown of the
Lifshitz scaling symmetry
\begin{equation}
t\to\lambda^z t,\qquad\vec x\to\lambda\vec x.
\end{equation}
In particular, the anomaly is given by
\begin{equation}
z\langle T_t^t\rangle+\langle T_i^i\rangle=\mathcal A,
\end{equation}
where $\mathcal A$ can be constructed out of geometrical invariants.  In contrast with
the relativistic case, non-relativistic scaling provides fewer constraints on the form of $\mathcal A$
\cite{Adam:2009gq,Gomes:2011di,Griffin:2011xs,Baggio:2011ha,Arav:2014goa}.
Moreover, the invariants that contribute have dimension $d+z$ (where $d$ is the number of
spatial dimensions) and may be formed out of a combination of time and space derivatives
with dimensions $z$ and $1$, respectively.  As a result, the structure of $\mathcal A$ depends
very much on the values of $z$ and $d$.

In the case of $z=2$ and $d=2$ Lifshitz, $\mathcal A$ is dimension four and has two possible
terms, with coefficients $C_1$ for a two time-derivative anomaly and $C_2$ for a four
space-derivative anomaly \cite{Baggio:2011ha,Arav:2014goa}.  A holographic calculation yields
\begin{equation}
\label{eq:4.2}
C_1 = \frac{1}{128\pi}\frac{2L}{G^{(4)}_N},\qquad C_2=0,
\end{equation}
which demonstrates that the Lifshitz radius $L$ is indeed directly related to the scale anomaly.
Similar results may be obtained for other values of $z$ and $d$.

We are, of course, more directly interested in the Schr\"odinger case, where there are additional
Galilean symmetries.  For $z=2$ Schr\"odinger, the Weyl anomaly was investigated in
\cite{Jensen:2014hqa}, and was shown to vanish for even-dimensional spacetimes (odd $d$).
For odd-dimensional spacetimes, the lowest derivative anomaly has the same structure as the
relativistic case in one dimension higher.  It would be of interest to more directly connect this
result with the radius $L$ that appears in Schr\"odinger holography.

\acknowledgments
We would like to thank Wenli Zhao for useful discussions. This work was supported in part by the US Department of Energy under grant DE-SC0007859.

\appendix
\section{Modified weak energy condition for Lifshitz spacetime}

In this appendix, we investigate the possibility of obtaining a holographic $c$-theorem from a modified weak energy condition in Lifshitz spacetime. We begin with the Lifshitz metric (\ref{eq:ABlif}), which we
repeat here for convenience
\begin{equation}
\label{eq:A1}
ds_{d+2}^2=-e^{2A(r)}dt^2+e^{2B(r)}d\vec x_d^2+dr^2,
\end{equation}
along with the definition (\ref{eq:ff}) of the flow functions $L(r)$ and $z(r)$.  The corresponding
Einstein tensor is
\begin{equation}
\label{eq:A2}
\begin{split}
G_{tt} &=\fft{g_{tt}}{L^2}d\left(-L'+\fft{d+1}2\right),\\
G_{ij} &= \frac{g_{ij}}{L^2}\left(z'L -(z+d-1)L'+ z^2 + (d-1)z +\frac{d(d-1)}{2}\right),\\
G_{rr} &=\frac{d}{L^2}\left(z+ \frac{d-1}{2}\right),
\end{split}
\end{equation}
and the Ricci scalar is
\begin{equation}
R = -\frac{2}{L^2}\left(z'L -(z+d)L'+z^2+dz+\frac{d(d+1)}{2}\right).
\end{equation}

The consequences of the null energy condition on $L(r)$ and $z(r)$ were investigated in
\cite{Liu:2012wf}, and the resulting inequalities are
\begin{equation}
L'\ge-(z-1),\qquad
z'\ge(z-1)(L'-d-z)/L.
\label{eq:necLif}
\end{equation}
When $z\ge1$, these inequalities may be combined to give (\ref{eq:Lifineq}).  In any case,
the null energy condition does not lead to monotonicity of $L(r)$.  In an attempt to obtain a
monotonic Lifshitz flow, we turn instead to the weak energy condition.

\subsection{Weak energy condition}

A conventional application of the weak energy condition is equivalent to the statement
\begin{equation}
\label{eq:A4}
G_{\mu \nu}t^\mu t^\nu \geq 0,
\end{equation}
for all future-directed time-like vectors $t^\mu$.  In this case, an upper bound for $L'$ is achieved
in the limit when $t^\mu$ approaches a null vector in the $t$--$x$ plane.  The result coincides with
the second inequality in (\ref{eq:necLif}), which may be expressed as
\begin{equation}
L' \leq z+d + \frac{z'L}{z-1}
\end{equation}
(assuming $z>1$).  On the other hand, a lower bound
\begin{equation}
\label{eq:A8}
L' \geq \frac{d+1}{2},
\end{equation}
achieved when $t^\mu$ is points purely at the time direction.

Note that this lower bound on $L'$ is incompatible with having a Lifshitz fixed point (where $L$ would
approach a constant).  This is actually not surprising, as the presence of a negative cosmological
constant, which can be expected in a Lifshitz background, can violate the weak energy condition.
For a fixed cosmological constant, it is possible to modify the weak energy condition to exclude
its contribution.  Of course, it is not always possible to disentangle the contribution of a constant
$\Lambda$ from a dynamical $\Lambda_{\rm eff}$.  Nevertheless, we investigate this possibility.

\subsection{Modified weak energy condition with an effective cosmological constant}

Since the lower bound on $L'$ given by \eqref{eq:A8} arises directly from the $G_{tt}$
in \eqref{eq:A2}, we may remove the $(d+1)/2$ contribution by imposing a subtracted
weak energy condition
\begin{equation}
(G_{\mu\nu}+\Lambda_{\rm eff}g_{\mu \nu})t^\mu t^\nu \geq 0.
\end{equation}
Choosing 
\begin{equation}
\Lambda_{\rm eff} = -\frac{d(d+1)}{2L^2}.
\end{equation}
then gives
\begin{equation}
\label{eq:A11}
0\leq L' \leq z+d + \frac{z'L}{z-1}.
\end{equation}
Note that $\Lambda_{\rm eff}$ is precisely the cosmological constant of pure AdS$_{d+2}$ with
radius $L$.

Although this subtracted weak energy condition allows for both Lifshitz fixed points and
monotonic flows for $L(r)$, it is not necessarily a well-defined energy condition on the matter
fields.  In particular, $\Lambda_{\rm eff}$ is implicitly defined through the flow function $L(r)$,
which in turn is obtained from the metric function $A(r)$, and not directly from the matter
sector.  We thus turn away from this possibility and consider another modification to the weak
energy condition that can be formulated more directly in terms of the stress tensor.

\subsection{Modified weak energy condition with a Ricci scalar}

Instead of an effective cosmological constant, we may add a geometric invariant
to the left-hand side of \eqref{eq:A4}.  An obvious choice would be to use the Ricci scalar, so
we consider a modification of the form
\begin{equation}
\label{eq:A16}
(G_{\mu \nu} + kg_{\mu \nu}R)t^\mu t^\nu \geq 0,
\end{equation}
where $k$ is a constant that we adjust to achieve $L' \geq 0$.

In Lifshitz spacetime with critical exponent $z$, choosing $k$ to be
\begin{equation}
k = \left(\frac{4z^2}{d(d+1)}+\frac{4z}{d+1}+2\right)^{-1},
\end{equation}
then gives rise to the inequality
\begin{equation}
0\leq L' \leq z+d,
\end{equation} 
where the lower bound is again achieved when $t$ points purely in the time direction, and the upper bound is achieved when $t$ approaches a null vector.
Note that this result is the same as \eqref{eq:A11} when $z$ is constant.

In fact, for $k$ to be a constant, we must take $z$ to be a constant as well.  Thus this
modified weak energy condition is only applicable to Lifshitz flows where $z$ is held fixed.
In this case, we can use the Einstein equation to rewrite \eqref{eq:A16} as a condition on
the stress tensor 
\begin{equation}
\label{eq:A19}
(T_{\mu \nu}- \frac{2k}{d-1}g_{\mu \nu}T)t^\mu t^\nu \geq 0.
\end{equation}
In order to better understand the meaning of this energy condition, we consider a perfect
fluid in Minkowski spacetime.  In this case, \eqref{eq:A19} gives two conditions on the
pressure $p$ and the density $\rho$
\begin{equation}
\rho + p  \geq 0, \qquad 
\left(1-\frac{2k}{d}\right)\rho +\left(\frac{6k}{d}\right)p \geq 0.
\end{equation}
This naturally limits to the usual weak energy condition
\begin{equation}
\rho + p \geq 0, \qquad
\rho  \geq 0,
\end{equation}
in the limit $k\to0$.

It is not entirely clear what the significance of such a modified weak energy condition is.
Moreover, many Lifshitz flows of interest would not necessarily be constrained to have
constant $z$.  So, in the end, the possibility of obtaining a holographic $c$-theorem for
Lifshitz spacetimes based on a physically well-motivated energy condition in the bulk
remains an open question.


\end{document}